\documentstyle[graphics]{aa}

\def\spose#1{\hbox to 0pt{#1\hss}} 
\def\lta{\mathrel{\spose{\lower 3pt\hbox{$\mathchar"218$}} 
        \raise 2.0pt\hbox{$\mathchar"13C$}}}      
\def\gta{\mathrel{\spose{\lower 3pt\hbox{$\mathchar"218$}} 
        \raise 2.0pt\hbox{$\mathchar"13E$}}} 
\def\msol{M$_\odot$}

\begin{document}

\title{No observational proof of the black-hole event-horizon}

\author{Marek A. Abramowicz\inst{1,3,4}, W{\l}odek Klu{\'z}niak\inst{2,3}, 
and Jean-Pierre Lasota\inst{3}}
\institute{
Department of Astronomy and Astrophysics, Chalmers University, 
412-96 G\"oteborg, Sweden\\
email : {\tt marek@fy.chalmers.se}
\and
Institute of Astronomy {\sl Johannes Kepler}, University of Zielona G\'ora,
Poland\\
email : {\tt wlodek@camk.edu.pl}
\and
Institut d'Astrophysique de Paris, 98bis Boulevard Arago, 75014 Paris, France\\
email : {\tt lasota@iap.fr}
\and
Universit\'e Pierre et Marie Curie, Paris 6
}

\date{Received / Accepted}
\titlerunning{No proof of event-horizon}
\authorrunning{Abramowicz, Klu{\'z}niak, Lasota}

\abstract{Recently, several ways of obtaining observational proof of the existence of
black-hole horizons have been proposed. We argue here that such proof
is fundamentally impossible: observations can provide arguments,
sometimes very strong ones, in favour of the existence of the event horizon, but they cannot
prove it. This applies also to future observations, which
will trace very accurately the details of the spacetime metric of a body suspected 
of being a black hole. 
}                             

\maketitle

\section{Introduction}

It is generally believed that the compact component in X-ray binary
systems is either a star possessing a material surface (a neutron star or
a quark star) or a black hole, i.e. an object whose surface is formed by
an event horizon. The evidence of the presence of a material surface is
obtained from two types of observations. First, stable periodic
pulsations of the X-ray emission indicate the presence of a strong
($10^{9} - 10^{15}$ G) magnetic field, which by the virtue of the
``no-hair" theorem excludes the presence of a black hole in the system.
Second, observations of X-ray bursts - thermonuclear explosions
occurring in matter accumulated at the surface of the compact object are
an obvious proof of the absence of an event horizon. 

Although X-ray pulsations or X-ray bursts indicate the presence of a
solid surface, their absence does not prove the presence of an event
horizon. ``Absence of evidence is not evidence of absence". However,
there is direct evidence that compact bodies in X-ray binaries form
(at least) two types of objects: their masses show a bimodal
distribution (see Miller, Shahbaz \& Nolan \cite{msn98}). Neutron star
masses are all concentrated around the ``canonical" value of 1.4~\msol\ 
whereas the second class of bodies, usually called ``black hole
candidates" have higher masses in the range of $\sim$ 5 to 18 \msol
(see e.g. Narayan, Garcia \& McClintock \cite{ngm01} and Greiner, Cuby
\& McCaughrean \cite{gcmc01}).  The reason for suspecting the more
massive bodies of being black holes is that their masses are higher
than the maximum mass of a neutron (or quark) star, which is never
larger than $\sim 3$ \msol\  (see e.g. Salgado et al.
\cite{setal94}).

In general, the maximum mass of a compact body can be expressed as
$8.4\left(\varepsilon_0/10^{14} {\rm g\ cm^{-3}}
\right)^{-1/2}$\msol, where $\varepsilon_0$ is the fiducial density
above which the equation of state is taken to be described by a
causality-limit equation of state (Rhoades \& Ruffini \cite{rr74};
Friedmann \& Ipser \cite{fi87}). Bahcall, Lynn \& Selipsky
(\cite{bls90}) showed that stars with a material surface can have masses
as high as 10 \msol, if one is willing to entertain configurations of
sub-nuclear density. These are the so-called ``Q-stars." The mean-field
description of nuclear interactions given by Bahcall et al.
(\cite{bls90}) allows baryonic matter to have densities this low.

Although, as shown by Miller et al. (\cite{msn98}), it is unlikely
that bodies with masses larger than 10 \msol\ are Q-stars---because
this would require unrealistically low densities at which hadronic
bulk matter would persist---``unlikely" is not a very satisfactory
argument in favour of the black-hole existence. One would rather wish
a ``positive'' proof of the event-horizon's existence. This has been
attempted by Ramesh Narayan and collaborators. The claim is that
properties of Advection Dominated Accretion Flows (ADAFs; Abramowicz
et al. \cite{aetal95}; Narayan \& Yi \cite{ny94},
\cite{ny95a},\cite{ny95b}) can be used to prove the existence of
event-horizons.

\section{Proof by ADAFs}

ADAFs describe accretion with very low radiative efficiency in which
energy released by viscous torques removing angular momentum from the
accreting matter is not radiated away but stored in the flow. If an ADAF
forms around a black hole, the stored energy will be lost forever under
the event horizon, whereas if the accreting body is a ``star" this
energy must be radiated away once matter lands on its surface.
Therefore, the argument runs, black holes should be dimmer than neutron
stars, quark stars, etc., {\sl if} in both cases an ADAF is present.

The best systems in which this hypothesis could be tested are the
so-called Soft X-ray Transients (SXTs) which are close binary systems
undergoing rare and powerful outbursts but spending most of their life
in a low luminosity quiescent state (see Tanaka \& Shibazaki
\cite{ts96} for a review). In SXTs, like in Low-Mass X-Ray Binaries
(LMXBs) in general, a compact body accretes matter lost by a Roche-lobe
filling low-mass stellar companion. The accreting matter forms a disc
whose instabilities trigger outbursts (see Lasota \cite{l01} for a
review of the instability model). Narayan, McClintock \& Yi
(\cite{nmy96}; see also Lasota, Narayan \& Yi \cite{lny96} and Narayan,
Barret \& McClintock \cite{nbm97}) proposed that quiescent SXT discs are
truncated and that the inner accretion flow forms an ADAF. This
hypothesis has been recently vindicated from the theoretical point of
view by Dubus, Hameury \& Lasota (\cite{dhl01}) and is supported by
observations (see Done \cite{cd02} for a review).

Narayan, Garcia \& McClintock (\cite{ngm97}) compared quiescent
luminosities of SXTs supposed to contain black holes with those of
neutron-star SXTs and realized that, in accordance with the prediction
of the ADAF model, systems containing black-hole ``candidates"
are dimmer. They came to the conclusion that they found evidence for the
presence of event horizons. 

This conclusion has been challenged by Chen et al. (\cite{chetal98})
who asserted that the relative dimness of black-hole candidate systems
was due solely to Narayan et al. (\cite{ngm97}) comparison method. Things
were clarified by Lasota \& Hameury (\cite{lh98}) who suggested comparing
systems with similar orbital period on the assumption such systems would
have similar accretion rates -- the ADAF model asserting only that accreting 
black holes should be dimmer than neutron stars for the {\sl same}
accretion rate. The new method showed, however, the same effect (Lasota
\& Hameury \cite{lh98}; Menou et al. \cite{metal99}), recently confirmed
by Garcia et al. (\cite{getal01}): black holes (candidates) are dimmer
than systems known to contain neutron stars, or at least stars with
surface.

This is a very strong argument in favour of the presence of event
horizons, in fact this is the most conservative conclusion. However, it 
is not a proof. 

\section{Arguments against evidence based on relative dimness of black
hole candidates}

The arguments against the claim that the relative dimness of black-hole
candidates is the proof of existence of event horizons are of two, not
unrelated, types. First, it has been argued that the accretion flow in
quiescent SXTs are not represented by ADAFs. 

Narayan \& Yi (\cite{ny95a}) and Blandford \& Begelman (\cite{bb99})
argued (see however Paczy\'nski \cite{bp98} and Abramowicz, Lasota \&
Igumenshchev \cite{ali00} for criticism of the argument) that ADAFs are
subject to mass loss and therefore the dimness of quiescent SXTs could
result from the low accretion rate onto the compact object - most of the
matter being lost with the wind. However, as shown by Menou et al.
(\cite{metal99}), such wind models do not offer an explanation of the
luminosity difference between neutron-star systems and those presumed to
contain black holes. In fact, these authors also pointed out that the
quiescent luminosity of neutron-star binaries is not consistent with the
assumption of a $\sim 10\%$ radiative efficiency. Since the attempt to
apply to these systems the windy-ADAF model of Quataert \& Narayan
(\cite{qn99}) failed, they proposed that the action of a magnetic
propeller could be answer. However, a compelling signature of this
effect has yet to be found.

Despite of this, Abramowicz \& Igumenshchev (2001) suggested that the
observed differences between quiescent luminosities of accreting black
holes and neutron stars is well explained by the occurrence in such
systems of a CDAF (Convection Dominated Accretion Flow; see Narayan,
Igumenshchev \& Abramowicz 2000) instead of an ADAF. They found that for
low viscosities accretion flows around compact bodies form ADAFs only in
their innermost regions but are convectively dominated at radii $R\gta
10^2R_{\rm S}$ (where $R_{\rm S}=2GM/c^2$ is the Schwarzschild radius).
In such flows emission comes mostly from the convective region; the
radiative efficiency is independent of accretion rate and equals
$\varepsilon_{\rm BH}= 10^{-3}$. Assuming that the efficiency of
accretion onto a neutron star is $\varepsilon_{\rm NS}\approx 0.1$ one
obtains the observed ratio between black-hole and neutron-star
luminosities. Unfortunately this cannot be the correct explanation
of the luminosity difference (Lasota \cite{l02}) because, as mentioned
above, neutron stars in quiescent transient systems do not seem to
accrete with a 0.1 efficiency.

Another class of argument asserts that  X-rays in quiescent SXTs  are 
not emitted by the accretion flow.

Brown, Bildsten, \& Rutledge (\cite{bbr98}) suggested that, in
neutron-star systems, most (or all) of the quiescent X-ray luminosity is
not due to accretion but results from cooling of the neutron-star crust
heated by nuclear reactions. This crust-cooling model does not seem to
be in perfect agreement with observations showing two spectral
components and a variable flux (see Rutledge et al. \cite{retal02} and
references therein). If the crustal-cooling model were right it would
imply different X-ray emission mechanisms for the two classes of
quiescent SXTs. However, luminosity variations observed also in
quiescent black-hole systems (see e.g. Garcia et al. \cite{getal01})
would rather suggest a common origin. Attempts to ascribe quiescent
X-ray luminosity in black-hole systems to active stellar companions
(Bildsten \& Rutledge \cite{br00}) are not based on a sound theoretical
foundation (Lasota \cite{l01}) and have been refuted by observations
(Garcia et al. \cite{getal01}). 

Menou (\cite{m02}) presented an argument based on the settling-flow
model of Medvedev \& Narayan (\cite{medna01}) in which the accretion
flow arrives with very low angular momentum at the surface of a rapidly
rotating compact object. The X-ray luminosity is then due to
rotation-energy loss by the accreting body. This requires viscous
contact between this body and the accreting matter. Menou (\cite{m02})
pointed out that if black-hole candidates had, contrary to neutron
stars, radii smaller than the inner-most stable orbit the accretion flow
would be supersonic and viscous contact impossible. Black-hole
candidates would be dimmer because unable to lose their rotational
energy.

Finally, we note that {\sl very compact} objects with a surface would be dimmer
than less compact objects, simply because of redshift and light bending.
If the surface is below the photon orbit, the fraction of ``outward moving''
photons which escape to infinity is in the Schwarzschild metric
\begin{equation}
{\Delta\Omega\over 2\pi}
=1-\left[1-{27\over4}{(1-R_S/R)\over( R/R_S)^2}\right]^{1/2}.
\end{equation} 
For the lowest possible value for a causality-limit equation of state
$R/R_S=9/8$, this factor and the redshift squared yield a luminosity at
infinity which is equal to only 0.040 of the luminosity at the source.

\section{Absence of X-ray bursts}

Three of the SXTs show millisecond pulsations, and two of them are X-ray
bursters. They all have very short orbital periods, 2 hr in the case of
SAX J1808.4-3658 (Wijnands and van der Klis \cite{wvdk}; Chakrabarty and
Morgan \cite{cm98}), 43.6 min for XTE J1751-305 (Markwardt et al.
\cite{marketal}), and 42 min for XTE J0929-314 (Galloway et al.
\cite{galetal}). It is perfectly well understood that occurrence of
coherent pulsations or of type I X-ray bursts is incompatible with the
presence of an event horizon, so none of these sources can be found on
the list of black hole candidates, even though their masses are unknown.

However, it is true, as pointed out by Narayan \& Heyl (\cite{nh02}), that
none of the longer (binary) period SXTs, with a measured mass function
greater than $3M_\odot$ is a type~I burster. Narayan \& Heyl (\cite{nh02})
compute instability of accretion onto a hypothetical $10M_\odot$ star
with a surface of radius between $(9/8)R_S$ and $3R_S$, and report that
for a range of accretion rates compatible with observations of X-ray
novae, the star is expected to give rise to an X-ray burst if the
accreted column density is $10^9\,{\rm g/cm^2}\le\Sigma\le 10^{11}\,{\rm
g/cm^2}$. From this, the authors conclude that black hole candidates
cannot have a surface, as they do not exhibit X-ray bursts. 

One concern is that the authors do not present the results separately
for the lowest column density considered, $10^9\,{\rm g/cm^2}$, and the
higher values $10^{10}\,{\rm g/cm^2}$ and $10^{11}\,{\rm g/cm^2}$---for
a $10M_\odot$ star with a $3R_S$ radius, the mass transferred in the
transient outburst $\sim 6\times10^{24}\,{\rm g/cm^2}$ corresponds to
$6\times10^9\,{\rm g/cm^2}$, so the X-ray burst expected at one of the
higher column densities may, in fact, not occur during a SXT outburst.
However, there is a more fundamental doubt as to the relevance of the
result.

Since the minimum radius of Q-star is 1.4 $R_S$ (Miller et al.
\cite{msn98}), Narayan \& Heyl (\cite{nh02}) consider not only objects
composed of matter whose properties have been described by Bahcall et
al. (\cite{bls90}) but also more compact configurations whose
microscopic properties are not known at all. Therefore there is no
reason to assume that the surface of such objects is composed of
ordinary matter and is in the temperature range required for X-ray
bursts to occur. The stellar surface could be too cold to support a
thermonuclear runaway. As a matter of fact, the accreted matter could
be converted right away to a more exotic form, as it would be on
contact with quark matter in the color-locked phase (Alford, Rajagopal
\& Wilczek \cite{arw}, Rapp et al. 1998), or with the skin of a
gravastar (Mazur \& Mottola \cite{MM01}, see below). This could happen
even at zero density, contrary to the hypothesis advanced by Narayan
\& Heyl (\cite{nh02}).  No nuclei, no bursts.

\section{Gravastars}

Mazur and Mottola (\cite{MM01}) have recently found a new static,
spherically symmetric, solution of Einstein's field equations. A
gravastar, as it is called, has the standard vacuum Schwarzschild
exterior, and an interior filled with matter that has the equation of
state $\rho = -p$. The interior is described by the de Sitter solution,
and is matched to the exterior vacuum solution in a very thin shell of
thickness on the order of the Planck length, $\lambda_P = 1.6 \times
10^{-33}$ cm. The gravastar has no horizon or singularity. Its rigid
surface is located at a radius just slightly greater than the
gravitational radius, $R_* = R_S + f\lambda_P$, $f \sim 2$. 

There are several purely theoretical objections that one could raise
against gravastars, none of them conclusive. For example, stellar-mass
gravastars have entropy smaller than ordinary stars with the same
masses and this would require extremely efficient cooling before
gravastars could form during stellar collapse. 

There is no {\it observational} way to distinguish what may seem to be a
Schwarzschild black-hole from a gravastar. To see this, let us denote
the surface redshift by 
\begin{equation}
 \varepsilon = \left ( 1 - {R_S \over R_*} \right)^{1/2} =
\left( {f\lambda_P \over R_*} \right)^{1/2}.
\end{equation} 
For astrophysically interesting gravastars, with mass greater than
$M_{\odot}$, i.e., $R_S > 3\times 10^{5}$ cm, this quantity is very
small, 
\begin{equation}
\varepsilon < 10^{-19} \ll 1.
\end{equation}

The power of any radiation emitted by the surface of a gravastar is
greatly reduced because only the radiation within the solid angle
$27\varepsilon^2/4$ around the normal to the surface escapes to
infinity. Further, because of gravitational redshift, the power of
radiation received by a distant observer is only $\varepsilon^2$ of what
was emitted at the gravastar's surface. Therefore, the power emitted
from the surface is reduced by 
\begin{equation}
\varepsilon^{4} < 10^{-75}
\end{equation}
by the time it reaches a distant observer. One should conclude that a
gravastar with mass greater than $M_{\odot}$ is to a distant observer 
as black as a black hole. 

\section{Conclusions}

We have shown that it is fundamentally impossible to give an
observational proof for the existence of a black-hole horizon. One could
argue that it is not necessary to give such a proof -- a black hole is a
specific space-time metric, whose properties can, in principle, be
determined through observations. If so, no `direct' proof would be 
necessary, assuming one could determine that the spacetime around a 
compact object corresponds to the Kerr solution of the Einstein
equations.

One way to distinguish a black hole from a rotating star is through the
study of orbital and other frequencies (e.g., epicyclic) of accreting
matter moving in strong-field gravity (Kato \cite{sk01}; Wagoner
\cite{wso}; Abramowicz \& Klu{\' z}niak \cite{ak01} \cite{ak02}).
Another method of determining the space-time geometry is by observation
of the energy spectrum reflected from an accretion disc deep in
the gravitational well of a compact object (Fabian et al. \cite{frsw},
\cite{firy}). Finally, the capture of stellar-mass compact objects by
supermassive black holes in galactic nuclei produces gravitational
radiation whose properties reflect the structure of black-hole space-time
(Ryan \cite{r95}; Hughes \cite{sc01}). Such gravitational radiation
could be observed by the gravitational-wave antenna LISA. 

The last method is the most powerful since it uncovers the compact
body's multipolar structure. Like the other methods, however, it suffers
from a fundamental weakness because it assumes that only a black hole can
be the `source' of the ($a\neq 0$) Kerr metric. Although it was shown
that it is very unlikely that other sources exist (Abramowicz, Lasota \&
Muchotrzeb \cite{alm}), such a possibility cannot be excluded.

Nevertheless, the case for the existence of black holes in the Universe 
is very strong and the evidence very convincing. We think, however, 
that a shadow of doubt will always cast its pall on our certainty in this 
matter. But it is a fertile doubt: it has already inspired new ideas and 
will surely continue to do so.


\listofobjects


\begin{thebibliography}{}


\bibitem[2001]{ak01} Abramowicz, M.A., Klu{\'z}niak, W.\ 2001, A\&A, 374, L19

\bibitem[2002]{ak02} Abramowicz, M.A., Klu{\'z}niak, W.\ 2002, astrop-ph/0203314

\bibitem[1976]{alm} Abramowicz, M.A., Lasota, J.-P., Muchotrzeb, B.\
1976, Comm. Math. Phys., 47, 109

\bibitem[2000]{ali00} 
Abramowicz, M.A., Lasota, J.-P., \& Igumenshchev, I.V.\ 2000, MNRAS, 314, 
775 

\bibitem[1995]{aetal95} Abramowicz, M.A., 
Chen, X., Kato, S., Lasota, J.-P., \& Regev,~O.\ 1995, ApJ, 438, L37

\bibitem[1998]{arw} Alford, M., Rajagopal, K., Wilczek, F.\ 1998, Physics
Letter B, 422, 247

\bibitem[1990]{bls90} Bahcall, 
S., Lynn, B.W., \& Selipsky, S.~B.\ 1990, ApJ, 362, 251

\bibitem[2000]{br00} Bildsten, L. \& 
Rutledge, R.E.\ 2000, ApJ, 541, 908 

\bibitem[1999]{bb99} Blandford, 
R.D. \& Begelman, M.C.\ 1999, MNRAS, 303, L1 

\bibitem[1998]{bbr98} Brown, 
E.F., Bildsten, L. \& Rutledge, R.E.\ 1998, ApJ, 504, L95 

\bibitem[1998]{cm98} Chakrabarty, 
D., Morgan, E.H.\ 1998, NATURE, 394, 346 

\bibitem[1998]{chetal98} Chen, W., Cui, W., Frank, 
J., King, A., Livio, M. \& Zhang,~S.\ 1998, in Accretion Processes in 
Astrophysical Systems: Some Like it Hot!, 347 

\bibitem[2002]{cd02} Done, C.\ 2002, Phil. Trans. RAS, submitted, astro-ph/0203246

\bibitem[2001]{dhl01} Dubus, G., 
Hameury, J.-M., \& Lasota, J.-P.\ 2001, A\&A, 373, 251 

\bibitem[1989]{frsw} Fabian, A.C., Rees, M.J., Stella, L., White, N.E., 1989, 
MNRAS, 238, 729

\bibitem[2000]{firy} Fabian, A.C., Iwasawa, K.,  Reynolds, C.S., 
Young, A.J., 2000, PASP, 324, 923

\bibitem[1987]{fi87} Friedman, J.L.~\& 
Ipser, J.R.\ 1987, ApJ, 314, 594

\bibitem[2002]{galetal} Galloway, D.K., Chakrabarty, D. Morgan, E.H., 
Remillard, R.A., \ 2002, ApJL, submitted, astro-ph/0206493

\bibitem[2001]{getal01} Garcia, M.R.,  
McClintock, J.E., Narayan, R., Callanan, P., Barret, D. \& Murray, S.S.
2001, ApJ, 553, L47

\bibitem[2001]{gcmc01} Greiner, J., 
Cuby, J.-G., \& McGaughrean, M.J.\ 2001, Nature, 414, 522 

\bibitem[2001]{sc01} Hughes, S.A., 2001, Class. Quantum. Grav. 18, 4067

\bibitem[2001]{sk01} Kato, S., 2001, PASJ, 53, 1

\bibitem[2001]{l01} Lasota, J.-P.\ 2001, New 
Astronomy Review, 45, 449

\bibitem[2002]{l02} Lasota, J.-P.\ 2002, in Proceedings of SF2A, EDPS 
Conference Series in Astronomy \& Astrophysics, 405

\bibitem[1998]{lh98} Lasota, J.-P. \& 
Hameury, J.-M.\ 1998, in Accretion Processes in Astrophysical Systems: Some 
Like it Hot!, 351

\bibitem[1996]{lny96} Lasota, J.-P., 
Narayan, R., \& Yi, I.\ 1996, A\&A, 314, 813

\bibitem[2002]{marketal} Markwardt, C.B.,  Swank, J.H., Strohmayer, T.E.,
 in't Zand, J.J.M., Marshall, F.E.\ 2002, ApJ, in press, astro-ph/0206491

\bibitem[2001]{MM01} Mazur, P.O., Mottola, E. 2001,  gr-qc/0109035

\bibitem[2001]{medna01} Medvedev, M.V. \& 
Narayan, R.\ 2001, ApJ, 554, 1255 

\bibitem[2001]{m02} Menou, K. 2001, in Proceedings of the 2nd KIAS 
Astrophysics  Workshop ``Current High-Energy Emission around Black Holes", 
astro-ph/0111469

\bibitem[1999]{metal99} Menou, K., Esin, A.A., 
Narayan, R., Garcia, M.R., Lasota, J.-P. \& McClintock, J.E.\ 1999, ApJ, 
520, 276 

\bibitem[1998]{msn98} Miller, 
J.C., Shahbaz, T., \& Nolan, L.A.\ 1998, MNRAS, 294, L25. 

\bibitem[2002]{nh02} Narayan, R., Heyl, J.S. 2001,  astro-ph/0203089

\bibitem[1994]{ny94} Narayan, R. \& Yi, I.\ 
1994, ApJ, 428, L13

\bibitem[1995a]{ny95a} Narayan, R. \& Yi, I.\ 
1995a, ApJ, 444, 231 

\bibitem[b]{ny95b} Narayan, R. \& Yi, I.\ 
1995b, ApJ, 452, 710

\bibitem[1997]{nbm97} 
Narayan, R., Barret, D., \& McClintock, J.E.\ 1997, ApJ, 482, 448

\bibitem[1997]{ngm97} 
Narayan, R., Garcia, M.R. \& McClintock, J.E.\ 1997, ApJ, 478, L79

\bibitem[2001]{ngm01} 
Narayan, R., Garcia, M.R. \& McClintock, J.E.\  2001, in Proc. IX Marcel 
Grossmann Meeting, eds. V. Gurzadyan, R.~Jantzen and 
R. Ruffini, Singapore: World Scientific, astro-ph/0107387

\bibitem[1996]{nmy96} Narayan, 
R., McClintock, J.E., \& Yi, I.\ 1996, ApJ, 457, 821

\bibitem[1998]{bp98} Paczy\'nski, B., 1998, Acta Astronomica, 
48, 667 

\bibitem[1999]{qn99} Quataert, E.~\& 
Narayan, R.\ 1999, ApJ, 520, 298

\bibitem[1998]{rssv} Rapp, R., Sch{\" a}fer, T., Shuryak, E., Velkovsky, M.\
1998, Phys. Rev. Letters, 81, 43

\bibitem[1974]{rr74} Rhoades, C.E., Jr., Ruffini, R. 1974, Phys. Rev. Lett., 32, 324

\bibitem[2002]{retal02} Rutledge, R.E., Bildsten, L., 
Brown, E.F., Pavlov, G.G. \& Zavlin, V.E., 2001, ApJ, 
in press, astro-ph/0204196

\bibitem[1995]{r95} Ryan, F.D., 1995, PRD, 52, 3159

\bibitem[1994]{setal94} Salgado, M., Bonazzola, S., 
Gourgoulhon, E., \& Haensel, P.\ 1994, A\&ASS, 108, 455

\bibitem[1996]{ts96} Tanaka, Y. \& 
Shibazaki, N.\ 1996, ARAA, 34, 607.

\bibitem[2001]{wso} Wagoner, R.V., Silbergleit, A.S., Ortega-Rodri{\' i}guez, 
M.\  2001, APJ, 559, L25

\bibitem[1998]{wvdk} Wijnands, 
R. \& van der Klis, M.\ 1998, NATURE, 394, 344  

\end{thebibliography}
\end{document}